\documentclass[11pt,oneside,fleqn]{article}

\usepackage[ansinew]{inputenc}
\usepackage[mathscr]{eucal}
\usepackage{amsmath,amssymb,amsthm}
\usepackage{graphicx}
\usepackage{cite}
\usepackage{hyperref}

\allowdisplaybreaks

\hypersetup{colorlinks, linkcolor=blue, citecolor=blue, urlcolor=blue}
\makeatletter
\long\def\@makecaption#1#2{%
  \vskip\abovecaptionskip\footnotesize
  \sbox\@tempboxa{#1. #2}%
  \ifdim \wd\@tempboxa >\hsize
    #1. #2\par
  \else
    \global \@minipagefalse
    \hb@xt@\hsize{\hfil\box\@tempboxa\hfil}%
  \fi
  \vskip\belowcaptionskip}
\makeatother

\newcommand{\p}{\partial}
\newcommand{\const}{\mathop{\rm const}\nolimits}

\newcommand{\todo}[1][\null]{\ensuremath{\clubsuit}}

\newcommand{\noprint}[1]{}

{\theoremstyle{definition}
\newtheorem{definition}{Definition}

\newtheorem*{remark*}{Remark}
}

\newcommand{\rsemioplus}{\mathbin{\mbox{$\lefteqn{\hspace{.76ex}\rule{.4pt}{1.2ex}}{\ni}$}}}

\newcommand{\checked}[1][\null]{\ensuremath{\boldsymbol{\surd}}}

\newcommand{\ve}{\varepsilon}

\begin{document}

\par\noindent {\LARGE\bf
The Korteweg--de Vries equation \\and its symmetry-preserving discretization
\par}

{\vspace{4mm}\par\noindent {Alexander Bihlo$^\dag$, Xavier Coiteux-Roy$^\ddag$ and Pavel Winternitz$^\S$
} \par\vspace{2mm}\par}

{\vspace{2mm}\par\noindent {\it
$^{\dag}$Department of Mathematics and Statistics, Memorial University of Newfoundland,\\
$\phantom{^\dag}$St.\ John's (NL), A1C 5S7, Canada
}}

{\vspace{2mm}\par\noindent {\it
$^{\ddag}$~D\'{e}partement de physique and Centre de recherches math\'{e}matiques, Universit\'{e} de Montr\'{e}al, C.P.\ 6128, succ.\ Centre-ville, Montr\'{e}al (QC) H3C 3J7, Canada
}}

{\vspace{2mm}\par\noindent {\it
$^{\S}$~D\'{e}partement de math\'{e}matiques et de statistique and Centre de recherches math\'{e}matiques, Universit\'{e} de Montr\'{e}al, C.P.\ 6128, succ.\ Centre-ville, Montr\'{e}al (QC) H3C 3J7, Canada
}}

{\vspace{2mm}\par\noindent {\it
$\phantom{^\dag}$~\textup{E-mail}:
abihlo@mun.ca, xavier.coiteux-roy@umontreal.ca, wintern@crm.umontreal.ca
}\par}

\vspace{4mm}\par\noindent\hspace*{8mm}\parbox{140mm}{\small
 The Korteweg--de Vries equation is one of the most important nonlinear evolution equations in the mathematical sciences. In this article invariant discretization schemes are constructed for this equation both in the Lagrangian and in the Eulerian form. We also propose invariant schemes that preserve the momentum. Numerical tests are carried out for all invariant discretization schemes and related to standard numerical schemes. We find that the invariant discretization schemes give generally the same level of accuracy as the standard schemes with the added benefit of preserving Galilean transformations which is demonstrated numerically as well.
}\par\vspace{2mm}

\section{Introduction}

This article is part of a general program the purpose of which is to study the possibility of discretizing the equations of physics while preserving their Lie point symmetries~\cite{doro91Ay,doro94Ay,doro11Ay,doro00By,doro00Ay,doro03Ay,doro04Ay,doro13Ay,levi91Ay, levi01Ay,levi06Ay,levi13Ay,levi14Ay,rodr04Ay,wint04Ay,wint13Ay}. There are both conceptual and practical reasons for doing this. From the conceptual point of view symmetries under rotations, Galilei or Lorentz transformations, conformal and other transformations are of primordial importance both in classical and quantum physics. It would be a pity to loose them when studying physical phenomena in a discrete world. From the practical point of view symmetries of differential equations determine many of the properties of solutions. Preserving symmetries in a discretization should provide difference systems that share some exact solutions with the original differential equations, or at least provide better approximations than noninvariant systems. In turn, this should have implications for numerical solutions. Thus, symmetry preserving discretizations should provide solutions that are in some sense ``better'' than ``standard'' discretizations.

The basic idea~\cite{doro11Ay,levi06Ay,wint04Ay} of this approach is to approximate a differential equation by a ``difference system'' consisting of several discrete equations. The solutions of this system determine the lattice and approximate the solution of the differential equation. In the continuous limit the solutions of the lattice equations reduce to identities (like $0=0$) and the remaining solutions go to the appropriate solution of the differential equation. The difference scheme is constructed out of invariants of the Lie point symmetry group $G$ of the differential equation. The action of $G$ on the independent and dependent variables is the same as for the continuous case and this action is assumed to be known. The action of $G$ is not prolonged to derivatives, but to all points of the lattice (the ``discrete jet space'').

This invariant discretization approach has been extensively applied to ordinary differential equation (ODEs). It has been shown that for first order ODEs an invariant discretization is exact~\cite{rodr04Ay}. The solution of an invariant difference scheme coincides point by point with the appropriate solution of the ODE. Moreover it is sufficient if the difference system is invariant under a one-dimensional subgroup of the symmetry group.

For second and third order ODEs it is often possible to integrate the invariant scheme directly and thus see explicitly how solutions of the difference scheme converge to those of the ODE~\cite{doro00By,doro04Ay,doro13Ay,wint13Ay}. It has been shown on the example of numerous second and third order nonlinear ODEs that the invariant discretizations provide more accurate numerical solutions than standard methods~\cite{bour06Ay,bour08Ay,rebe09Ay}. This is specially so in the neighborhood of singularities where invariant methods, as opposed to standard ones, make it possible to continue solutions beyond the singularities.

For partial differential equations (PDEs) the first application of Lie group theory to numerical methods is, to our knowledge, due to Shokin and Yanenko~\cite{shok83Ay,yane68Ay}. Their approach ``Differential approximation'' is quite different from ours (for a comparison see~\cite{levi13Ay}).

Quite a few articles devoted to the symmetry adapted discretization of PDEs have appeared over the last 20 years (see e.g.~\cite{baki97Ay,bihl12Cy,bihl13By,bihl12By,budd01Ay,budd98Ay, budd99Ay,doro91Ay,doro00Ay,doro03Ay,kim04Ay,kim08Ay,levi01Ay,olve01Ay,rebe11Ay,rebe14Ay,vali05Ay}). Invariant discretizations of the Korteweg--de Vries (KdV) equation were presented in~\cite{doro94Ay,doro11Ay,vali05Ay}.

The purpose of this article is to study invariant discretizations of the KdV equations in greater depth. Thus we will compare the known invariant discretizations amongst each other and propose new ones. All of them will be tested as numerical schemes for known exact solutions. Their accuracy and stability will be evaluated by comparing with known analytic solutions.

The KdV equation is very suitable for such a study. On one hand, it is an integrable equation so a very large body of analytical solutions is known (due to inverse scattering techniques~\cite{ablo91Ay,gard67Ay}). On the other hand the KdV equation has an interesting Lie point symmetry group that includes Galilei invariance. It is a prototype of a Galilei invariant evolution equation that can be invariantly discretized on a mesh with horizontal time lines, but not on an orthogonal one (nor any other equally spaced one).

The original invariant discretizations~\cite{doro94Ay,doro91Ay} essentially correspond to using the Lagrange formulation of hydrodynamics in the continuous limit. We suggest an alternative discretization that is natural in the Eulerian formalism, especially when combined with adaptive computational schemes.

In Section~\ref{sec:ContinuousKdV} we review some well known results on the symmetry group of the continuous KdV equation and on its known analytical solutions. We also present the Lagrangian form of the KdV equation. The invariant discretizations are presented in Section~\ref{sec:DiscretizationKdV}. All numerical results are concentrated in Section~\ref{sec:ComputationalResultsKdV}. The final Section~\ref{sec:ConclusionKdV} is devoted to the conclusions.

\section{The continuous KdV equation}\label{sec:ContinuousKdV}

We shall write the KdV equation in the form
\begin{equation}\label{eq:KdV}
 u_t + uu_x + u_{xxx}=0.
\end{equation}
Its Lie point symmetry group is well-known (see e.g.~\cite{olve86Ay}). A basis for its Lie algebra~$\mathfrak g$ is given by the vector fields
\begin{equation}
 \mathcal D= 3t\p_t+x\p_x-2u\p_u,\quad \mathcal B=t\p_x+\p_u,\quad \mathcal P_1=\p_x,\quad \mathcal P_0=\p_t,
\end{equation}
corresponding to dilations, Galilei boosts and space and time translations, respectively.

The symmetry algebra $\mathfrak g$ has precisely five conjugacy classes of one-dimensional subalgebras. A representative list of these classes is given by the algebras
\begin{equation}\label{eq:InequivalentSubalgebrasKdV}
 \{\mathcal D\},\quad \{\mathcal B\},\quad \{\mathcal B+\mathcal P_0\},\quad \{\mathcal P_0\},\quad \{\mathcal P_1\}.
\end{equation}
Conjugacy is considered under the group of inner automorphisms of~\eqref{eq:KdV}, extended by the simultaneous reflections of $x$ and $t$
\begin{equation}
 \mathrm{R}x=-x,\quad \mathrm{R}t=-t,\quad \mathrm{R}u=u.
\end{equation}
Thus, $G=\mathrm{R}\rsemioplus G_0$, where $G_0=e^{d \mathcal D}e^{v \mathcal B}e^{t_0\mathcal P_0}e^{x_0\mathcal P_1}$.

The group can be used to get new solutions from known ones. If $u(t,x)$ is a solution of the KdV equation then so are $u(-t,-x)$ and
\begin{equation}\label{eq:GroupTransformationsKdV}
 \tilde u(\tilde t,\tilde x)=e^{2d}u\left(e^{-3d}(t-t_0),e^{-d}(x-x_0-v(t-t_0))\right)+e^{-2d}v,\qquad d,v,t_0,x_0\in\mathbb{R},
\end{equation}
where $d$, $v$, $t_0$ and $x_0$ are group parameters.

\subsection{Lagrangian formulation of the KdV equation}

The original form of the KdV equation~\eqref{eq:KdV} is written in \emph{Eulerian variables}, i.e.\ the velocity $u$ is a function of time and space, $u=u(t,x)$. An alternative to the Eulerian form is the \emph{Lagrangian form}. In the Lagrangian description of fluid mechanics the velocity $u$ is a function of time and of the original position of the \emph{fluid particle} $\xi$. Assuming that the fluid particles maintain their identity (hence $\xi$ is independent of time), one needs to express the KdV equation as an equation for $u=u(\tau,x(\tau,\xi))$, where $\tau=t$. Using the chain rule, the Eulerian form of the KdV equation~\eqref{eq:KdV} is transformed to
\begin{equation}\label{eq:KdVComputationalCoordinates}
 u_\tau +(u-x_\tau)\frac{u_\xi}{x_\xi}+ \frac{1}{x_\xi}\left(\frac{1}{x_\xi}\left(\frac{u_\xi}{x_\xi}\right)_\xi\right)_\xi=0.
\end{equation}
Up to now, no particular relation between the original physical coordinate $x$ and the new Lagrangian coordinate $\xi$ has been imposed. In the classical Lagrangian framework, this change of coordinates is specified by setting
\begin{equation}\label{eq:RelationLagrangianEulerian}
 x_\tau(\tau,\xi)=u(x(\tau,\xi),t).
\end{equation}
In other words, the change of variables from the Lagrangian coordinates to the Eulerian coordinates is completed upon integrating the equation for the particle trajectories~\eqref{eq:RelationLagrangianEulerian}. The KdV equation in Lagrangian coordinates then reduces to
\[
  u_\tau + \frac{1}{x_\xi}\left(\frac{1}{x_\xi}\left(\frac{u_\xi}{x_\xi}\right)_\xi\right)_\xi=0.
\]

The change of coordinates from the Eulerian form~\eqref{eq:KdV} to the form~\eqref{eq:KdVComputationalCoordinates} is more general than the particular Lagrangian case given through~\eqref{eq:RelationLagrangianEulerian}. In the more general case, the variables $(\tau,\xi)$ are referred to as the \emph{computational} coordinates. From the numerical point of view, using the KdV equation in computational coordinates~\eqref{eq:KdVComputationalCoordinates} gives the perspective of defining the relation $x=x(\tau,\xi)$ in such a manner that the evolution of the discretization grid is coupled to the evolution of the KdV equation itself. This is the main idea of using adaptive numerical schemes~\cite{huan10By}. The importance of such schemes in the framework of invariant discretization will be clarified in Section~\ref{sec:DiscretizationKdV}.

We should like to stress here that even for the more general form~\eqref{eq:KdVComputationalCoordinates} of the KdV equation with yet unspecified relation $x=x(\tau,\xi)$ it makes sense to fix the transformation $\tau=t$. This guarantees that the resulting equation will be of evolutionary type (though it would be sufficient to put $\tau=\tau(t)$).

\subsection{Symmetry reduction and exact solutions}\label{sec:SymmetryReduction}

One of the reasons why exact analytical solutions of PDEs are useful is that they can be used to check the accuracy of numerical algorithms, in particular the invariant discretizations to be presented below. For integrable equations with nontrivial symmetry groups (like the KdV equation) there exist two main sources of exact solutions. One is symmetry reduction, producing solutions invariant under some subgroup of the symmetry group. The other is the method of inverse scattering and its generalizations that lead to multisoliton and periodic and quasiperiodic solutions.

Let us start with the method of symmetry reduction. In order to reduce the KdV equation to an ODE we impose that the solution $u(t,x)$ be invariant under a one-dimensional subgroup $of G_0$ corresponding to a one-dimensional subalgebra of the symmetry algebra $\mathfrak g$. The classification of these subalgebras leads to the list~\eqref{eq:InequivalentSubalgebrasKdV}. Invariance under a subgroup corresponding to the algebra element $X=\tau\p_t+\zeta\p_x+\phi\p_u$ corresponds to imposing that $u(t,x)$ in addition to~\eqref{eq:KdV} should satisfy the quasilinear first order PDE
\begin{equation}\label{eq:InvariantAnsatz}
 \tau u_t+\zeta u_x=\phi.
\end{equation}
This equation is solved and the result is put into the KdV equation~\eqref{eq:KdV} which reduces to an ODE.

Let us run through the individual subalgbreas listed in~\eqref{eq:InequivalentSubalgebrasKdV}.

\medskip

\noindent\textbf{(i)} $\boldsymbol{\mathcal P_1=\p_x.}$ From~\eqref{eq:InvariantAnsatz} we obtain $u=f(t)$ and~\eqref{eq:KdV} implies that
\begin{equation}\label{eq:ConstantSolution}
 u(t,x)=A.
\end{equation}
Thus, the only solution of the KdV invariant under space translations is a constant.

\medskip

\noindent\textbf{(ii)} $\boldsymbol{\mathcal B=t\p_x+\p_u.}$ From~\eqref{eq:InvariantAnsatz} we get the reduction formula
\[
 u(t,x)=\frac{x}{t}+f(t).
\]
Substituting into~\eqref{eq:KdV} and solving the obtained ODE for $f(t)$ we find $f(t)=\frac{A}{t}$. Applying the group transformations~\eqref{eq:GroupTransformationsKdV} we obtain the Galilei (and dilation) invariant solution
\begin{equation}\label{eq:GalileiInvariantSolution}
 u(t,x)=\frac{x-x_0}{t-t_0}.
\end{equation}

\noindent\textbf{(ii)} $\boldsymbol{\mathcal B+\mathcal P_0=t\p_x+\p_u+\p_t.}$ The reduction formula following from~\eqref{eq:InvariantAnsatz} is
\[
 u(t,x)=t+f(\gamma),\quad \gamma=x-\frac12t^2.
\]
The KdV equation reduces to
$
 f'''+ff'+1=0.
$
Integrating once and putting
\[
 f(\gamma)=-12^{3/5}P\left[\left(\frac{1}{12}\right)^{1/5}(\gamma)+\delta\right]
\]
we find that $P(z)$ satisfies the first Painlev\'{e} equation
\begin{equation}\label{eq:ReductionKdVPainleveI}
 P''=6P^2+z,
\end{equation}
see~\cite{grom02Ay,ince56Ay}. The corresponding solution of the KdV equation is
\[
 u(x,t)=t-12^{3/5}P_{\rm I}\left[\left(\frac{1}{12}\right)^{1/5}(\gamma)+\delta\right],
\]
where $P_{\rm I}$ is the first Painlev\'{e} transcendent and $\delta$ is an arbitrary constant. No elementary solutions of~\eqref{eq:ReductionKdVPainleveI} are known.

\medskip

\noindent\textbf{(iv)} $\boldsymbol{\mathcal D=3t\p_t+x\p_x-2u\p_u.}$ The reduction formula~\eqref{eq:InvariantAnsatz} in this case yields
\[
 u=t^{-2/3}F(\gamma),\quad \gamma=xt^{-1/3},
\]
where $F(\gamma)$ satisfies
\begin{equation}\label{eq:ReductionKdVPainleveIIAnsatz}
 F'''+FF'-\frac13\gamma F'-\frac23F=0.
\end{equation}
The Miura transformation~\cite{olve86Ay}
$
 F=w'-w^2/6
$
and subsequent integration takes~\eqref{eq:ReductionKdVPainleveIIAnsatz} into
\begin{equation}\label{eq:ReductionKdVPainleveIIAnsatz2}
 w_{\gamma\gamma}=\frac{1}{18}w^3+\frac13\gamma w+k.
\end{equation}
Eq.~\eqref{eq:ReductionKdVPainleveIIAnsatz2} can be reduced to the equation
\begin{equation}\label{eq:ReductionKdVPainleveIIAnsatz3}
 P''=2P^3+zP+\alpha,
\end{equation}
where $\alpha$ is an arbitrary constant. This is the equation for the second Painlev\'{e} transcendent $P_{\rm II}$. Finally, the dilationally invariant solution of the KdV equation is
\begin{equation}
 u_\alpha(t,x)=2 (3)^{1/3}t^{-2/3}(P''_\alpha(\gamma)-P^2_\alpha(\gamma)),\quad \gamma=xt^{-1/3},
\end{equation}
where $P_\alpha$ is a solution of~\eqref{eq:ReductionKdVPainleveIIAnsatz3}.
Contrary to $P_{\rm I}$, the $P_{\rm II}$ equation allows two families of elementary solutions for special values of the parameter~$\alpha$~\cite{grom02Ay}. For integer values~$\alpha=\pm n$ these are rational solutions. For half integer values $\alpha=\pm(2n+1)/2$ the solutions are expressed in terms of Airy functions. In both cases they satisfy $P_\alpha=-P_{-\alpha}$ and are listed in~\cite{grom02Ay} for low values of $n$. For the combination $W_\alpha=P''_\alpha-P^2_\alpha$ we observe an additional relation, namely $W_{n+1}=-W_{-n}$, so for convenience we restrict to $\alpha=0,-1,-2,-3$. We thus obtain the following dilationally invariant solutions of the KdV equation
\begin{align}\label{eq:DilationInvariantSolutionsKdV}
\begin{split}
 &u_0=0,\quad, u_{-1}=-\frac{12}{x^2},\quad u_{-2}=-\frac{36x(24t-x^3)}{(12t+x^3)^2},\\
 &u_{-3}=-\frac{72(x^9+5400x^3t^2+43200t^2)x}{(720t^2-60x^3t-x^6)^2}.
\end{split}
\end{align}
The solution $u_0$ is also invariant under space and time translations, $u_{-1}$ is also invariant under time translations.
\medskip

\noindent\textbf{(v)} $\boldsymbol{\mathcal P_0=\p_t.}$ Solutions invariant under time translations have the form $u=f(x)$. A Galilei transformations boosts such a solution to a traveling wave $u=f(x-\lambda t)+\lambda$.

Substituting into the KdV equation and integrating twice we get an ODE that can be written as
\begin{equation}\label{eq:ReductionKdVStationary}
 (f')^2=-\frac13(f-a)(f-b)(f-c),\quad a+b+c=0.
\end{equation}
The roots of the polynomial in~\eqref{eq:ReductionKdVStationary} can all be real. Then we order them to have $a\ge b\ge c$. The other possibility is $a\in\mathbb{R}$, $b=\bar c=p+iq$, with $q>0$, $p,q\in\mathbb{R}$.

We are interested in real solutions only. They may be finite or singular (for $x\in\mathbb{R}$), periodic or localized. Let us run through the individual cases. Solutions are expressed in terms of Jacobi elliptic functions~\cite{byrd71Ay} or degenerate cases thereof.

\smallskip

\noindent {\it Cnoidal waves:} $c<b\le f\le a$, $b< a$. The solution in this case reads
\begin{equation}\label{eq:CnoidalWave}
 u(t,x)=b+(a-b)\mathrm{cn}^2(\omega x,k),\quad k=\sqrt{\frac{a-b}{2a+b}},\quad \omega=\sqrt{\frac{2a+b}{3}},\quad 2a+b>0.
\end{equation}
We can apply a Galilei boost with $v=-b$ and obtain the more usual form
\begin{equation}\label{eq:CnoidalWaveBoosted}
 u(t,x)=(a+v)\mathrm{cn}^2(\omega (x-vt),k),\quad k=\sqrt{\frac{a+v}{2a-v}},\quad \omega=\sqrt{\frac{2a-v}{3}}.
\end{equation}
\smallskip

\noindent {\it Soliton:} $c=b\le f\le a$,\quad $b=-\frac{a}{2}$, $k=1$, $\omega=\frac12\sqrt{\frac{a}{2}}$. The associated solution of the KdV equation is
\begin{equation}\label{eq:SolitonSolution}
 u(t,x)=-\frac{a}{2}+\frac{3a}{2}\frac{1}{\mathrm{cosh}^2\frac{1}{2}\sqrt{\frac{a}{2}}x},\quad a>0.
\end{equation}
After a boost with $a=2v$ we have the usual KdV soliton
\begin{equation}\label{eq:SolitonSolutionBoosted}
u(t,x)=\frac{3v}{\mathrm{cosh}^2\frac{1}{2}\sqrt{v}(x-vt)}.
\end{equation}
\smallskip

\noindent {\it Singular snoidal solution:} $f\le c< b<a$. The solution of the KdV equation reads
\begin{equation}\label{eq:SnoidalSolution}
 u(t,x)=a-\frac{a-c}{\mathrm{sn}^2(\omega x,k)},\quad \omega=\frac12\sqrt{\frac{a-c}{3}}, \quad k=\sqrt{\frac{2a+c}{a-c}}.
\end{equation}
\smallskip

\noindent {\it Singular soliton:} $f\le c=b< a$. The solution of the KdV equation in this case is
\begin{equation}\label{eq:SingularSolitonSolution}
 u=-\frac{a}{2}\left(1+\frac{3}{\sinh^2(\omega x)}\right),\quad \omega=\frac12\sqrt{\frac{a}{2}}.
\end{equation}
\smallskip

\noindent {\it Singular trigonometric solution:} $f\le c < b =a$. We obtain the solution
\begin{equation}\label{eq:SingularTrigonometricSolution}
 u=a-\frac{3a}{\sin^2(\omega x)},\quad \omega=\frac12\sqrt{a}.
\end{equation}
\smallskip

\noindent {\it Singular algebraic soliton:} $a=b=c=0$. The solution of the KdV equation in this case reduces to
\begin{equation}\label{eq:SingularAlgebraicSolitonSolution}
 u=-\frac{12}{x^2},
\end{equation}
which coincides with the solution $u_{-1}$ listed in~\eqref{eq:DilationInvariantSolutionsKdV} which is thus invariant under dilations and time translations. Galilei transformations take it into
\begin{equation}\label{eq:SingularAlgebraicSolitonSolutionBoosted}
 u(t,x)=-\frac{12}{(x-vt)^2}+v.
\end{equation}
\smallskip

\noindent {\it Real solutions corresponding to complex roots:} $f\le a\in \mathbb{R}$, $b=-\frac{a}{2}+iq$,\quad $c=-\frac{a}{2}-iq$, $q>0$. The corresponding solution of the KdV equation is
\begin{equation}\label{eq:RealSolutionComplexRoot}
 u(t,x)=a-A\frac{1+\mathrm{cn}(\omega x,k)}{1-\mathrm{cn}(\omega x,k)}, \quad A=\sqrt{\frac{9a^2}{4}+q^2},\quad \omega=\sqrt{\frac{A}{3}},\quad k^2=\frac{(A+\frac{3a}{2})^2+q^2}{4A^2}.
\end{equation}
An elementary special case is obtained for $k=1$, i.e.\ $a=\pm\frac23$, $A=\sqrt{1+q^2}$, namely
\begin{equation}\label{eq:RealSolutionComplexRootSpecialCase}
 u(t,x)=\pm\frac23-\sqrt{1+q^2}-\frac{\sqrt{1+q^2}}{\sinh^2\frac{\omega x}{2}},\quad \omega=\sqrt{\frac{1+q^2}{3}}.
\end{equation}

Other exact solutions are obtained by the inverse scattering method~\cite{ablo91Ay,gard67Ay}. Amonst them the most relevant for this article is the double soliton
\begin{align}\label{eq:DoubleSolitonSolution}
\begin{split}
&u(t,x)=12\frac{\partial^2}{\partial x^2}\ln(1+B_1e^{iQ_1}+B_2e^{iQ_2}+AB_1B_2e^{i(Q_1+Q_2)}),\\
&Q_1=a_1x-a_1^3t,\quad a_2x-a_2^3t,\quad A=\left(\frac{a_1-a_2}{a_1+a_2}\right)^2,
\end{split}
\end{align}
where $a_1$, $a_2$, $B_1$ and $B_2$ are arbitrary constants. Real solutions are obtained by putting $a_1=i\alpha_1$, $a_2=i\alpha_2$ with $\alpha_1,\alpha_2,B_1,B_2\in\mathbb{R}$.

Many other solutions ($n$-soliton, multigap quasiperiodic solutions, etc.) are available in the literature~\cite{dubr75Ay,dubr76Ay,grom02Ay,kric80Ay,poly04Ay}.

\section{Invariant discretization of the KdV equation}\label{sec:DiscretizationKdV}

\subsection{Invariant discretization on a ten point stencil}

The KdV equation is a scalar (1+1)-dimensional evolution equation. In the finite difference approximation on the $t$-$x$-plane, the continuous space of independent variables $(t,x)$ is sampled by a collection of finite points $\{P^n_i\}$ only. Here and in the following, we use the double index notation $(t^n_i,x^n_i)$ to denote a discrete point in this $t$-$x$-plane, where $i\in\mathbb{Z}$ is the spatial index and $n\in\mathbb{N}$ is the temporal index. Likewise, the dependent functions are defined on the associated points $\{P^n_i\}$ only, i.e.\ $u^n_i=u(t^n_i,x^n_i)$.

A partial differential equation $\mathcal L\colon \Delta (x,u^{(q)})=0$, where $u^{(q)}$ denotes all the derivatives of $u$ with respect to $t$ and $x$ up to order $q$, is discretized in a symmetry-preserving manner if it is expressed by a consistent finite difference approximation that can be written as a function of the finite difference invariants of the symmetry group of the equation itself. By \emph{consistent} it is meant that in the continuous limit (i.e.\ the distance between the points $\{P^n_i\}$ goes to zero) the finite difference approximation converges to the original differential equation~$\mathcal L$.

In writing this discretization, it is not only necessary to define a finite difference approximation of the differential equation~$\mathcal L$ itself but also to specify the lattice of points $\{P^n_i\}$ in an invariant fashion. In other words, the equation $\mathcal L$ is replaced by a system of finite difference equations of the form
\[
 \Delta S\colon E_\alpha(t^n_i,x^n_i,u^n_i)=0,\quad \alpha=1,\dots,N,\quad  i_{\rm min}\le i \le i_{\rm max},\quad  0\le n \le n_{\rm max}
\]
where the number of equations $N$ in the system $\Delta S$ is at least $N=3$.

The general method for finding invariant numerical schemes using difference invariants can be found e.g.\ in~\cite{doro11Ay,levi06Ay}. Here we only present the respective computations for the KdV equation. We should also like to mention here that there is another method for finding invariant discretization schemes that rests on invariantization using \emph{equivariant moving frames}. For more information on this alternative method, see e.g.~\cite{bihl12Cy,kim08Ay,olve01Ay,rebe11Ay}.

The minimum number of points in the stencil to discretize the derivatives in the KdV equation is five as spatial derivatives up to order three and a first order time derivative have to be approximated. In order to increase the accuracy of the finite difference approximation we introduce an extended ten point stencil. Lower order approximations can be obtained by restricting oneself to a subset of these 10 stencil points.

\begin{figure}[!ht]\label{fig:Stencil10Points}
 \centering
 \includegraphics[scale=0.7]{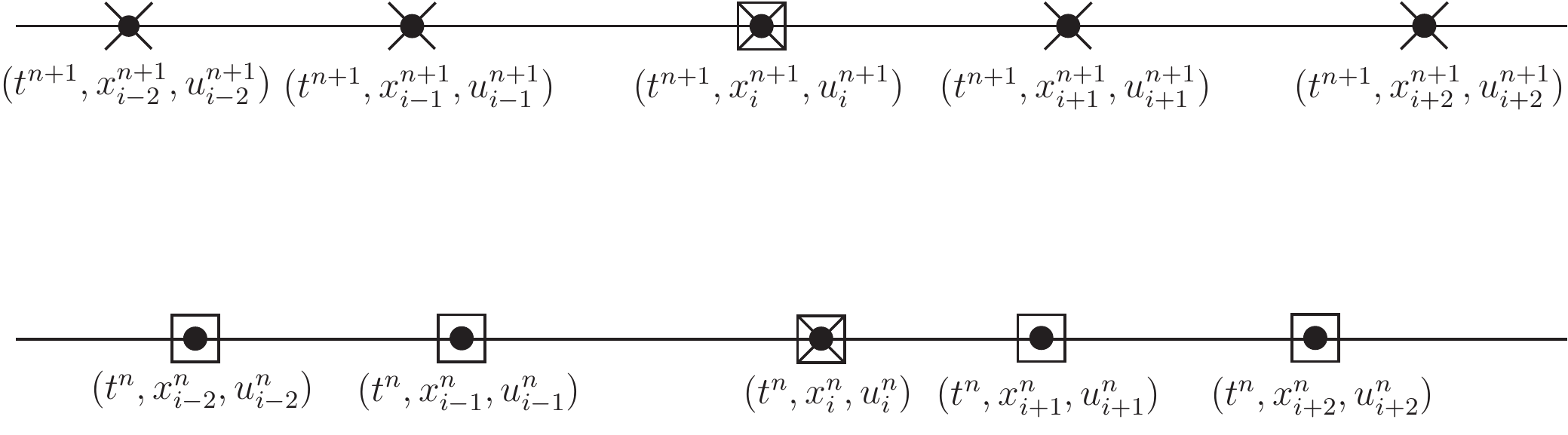}
 \caption{Stencils for the discretization of the KdV equation: Ten point stencil (solid circles). Explicit six point stencil (squares). Implicit six point stencil (crosses).}
\end{figure}

The stencils used are depicted in Fig.~\ref{fig:Stencil10Points}. It can be seen that a two-step time integration is employed allowing for either forward Euler (six point stencil, squares), backward Euler (six point stencil, crosses) or trapezoidal time integrators (ten point stencil, solid circles). Invariant numerical schemes using higher order time-stepping are possible as well but will not be presented here.

To simplify the notation, we also introduce the following abbreviations
\begin{align*}
 &\Delta \tau = t^{n+1}-t^n,\qquad h^n_i=x^n_{i+1}-x^n_i,\qquad
 Du_i^n=\frac{u^n_{i+1}-u^n_i}{h_i^n},
\end{align*}
for the spacings and elementary first order discrete derivatives. Note that the spacing in time does not carry an index as we use equally spaced, horizontal time layers only. It is readily checked that variable time-stepping would leave the following numerical scheme invariant as well, as long as the time-step control is invariant itself. See the similar discussion for the spatial adaptation strategies presented in Section~\ref{sec:AdaptiveDiscretizationKdV}.

The prolongation of vector fields of the maximal Lie invariance algebra $\mathfrak g$ to the stencil shown in Fig.~\ref{fig:Stencil10Points} yields
\begin{align}\label{eq:ProlongationOfVectorFieldsKdV}
\begin{split}
 &\p_{t^n}+\p_{t^{n+1}},\\
 &\p_{x^n_{i}}+\p_{x^n_{i+1}}+\p_{x^n_{i+2}}+\p_{x^n_{i-1}}+\p_{x^n_{i-2}}+ \p_{x^{n+1}_{i}}+\p_{x^{n+1}_{i+1}}+\p_{x^{n+1}_{i+2}}+\p_{x^{n+1}_{i-1}}+\p_{x^{n+1}_{i-2}},\\
 &t^n(\p_{x^{n}_{i}}+\p_{x^{n}_{i+1}}+\p_{x^{n}_{i+2}}+\p_{x^{n}_{i-1}}+\p_{x^{n}_{i-2}})+t^{n+1}(\p_{x^{n+1}_{i}}+\p_{x^{n+1}_{i+1}}+\p_{x^{n+1}_{i+2}}+\p_{x^{n+1}_{i-1}}+\p_{x^{n+1}_{i-2}})\\
 &\quad+\p_{u^{n}_{i}}+\p_{u^{n}_{i+1}}+\p_{u^{n}_{i+2}}+\p_{u^{n}_{i-1}}+\p_{u^{n}_{i-2}}+ \p_{u^{n+1}_{i}}+\p_{u^{n+1}_{i+1}}+\p_{u^{n+1}_{i+2}}+\p_{u^{n+1}_{i-1}}+\p_{u^{n+1}_{i-2}},\\
 &x^{n+1}_{i}\p_{x^{n+1}_{i}}+x^{n+1}_{i+1}\p_{x^{n+1}_{i+1}}+x^{n+1}_{i+2}\p_{x^{n+1}_{i+2}}+ x^{n+1}_{i-1}\p_{x^{n+1}_{i-1}}+x^{n+1}_{i-2}\p_{x^{n+1}_{i-2}}\\
 &\quad+3(t^{n+1}_{i}\p_{t^{n+1}_{i}}+t^{n+1}_{i+1}\p_{t^{n+1}_{i+1}}+t^{n+1}_{i+2}\p_{t^{n+1}_{i+2}}+ t^{n+1}_{i-1}\p_{t^{n+1}_{i-1}}+t^{n+1}_{i-2}\p_{t^{n+1}_{i-2}})\\
 &\quad-2(u^{n+1}_{i}\p_{u^{n+1}_{i}}+u^{n+1}_{i+1}\p_{u^{n+1}_{i+1}}+u^{n+1}_{i+2}\p_{u^{n+1}_{i+2}}+ u^{n+1}_{i-1}\p_{u^{n+1}_{i-1}}+u^{n+1}_{i-2}\p_{u^{n+1}_{i-2}}).
\end{split}
\end{align}

A complete list of functionally independent finite difference invariants annihilated by the prolonged infinitesimal generators on the ten point stencil~\eqref{eq:ProlongationOfVectorFieldsKdV} is exhausted by
\begin{align}\label{eq:DifferenceInvariants10points}
\begin{split}
 &I_1=\frac{h^{n}_{i-1}}{h^{n}_{i}},\quad
 I_2=\frac{h^{n}_{i+1}}{h^{n}_{i}},\quad
 I_3=\frac{h^{n}_{i-2}}{h^{n}_{i}},\quad
 I_4=\frac{h^{n+1}_{i}}{h^{n}_{i}},\quad
 I_5=\frac{h^{n+1}_{i-1}}{h^{n}_{i}},\quad
 I_6=\frac{h^{n+1}_{i+1}}{h^{n}_{i}},\\
 &I_7=\frac{h^{n+1}_{i-2}}{h^{n}_{i}},\quad
 I_8=\frac{(h^n_i)^3}{\Delta\tau},\quad
 I_9=\frac{x^{n+1}_i-x^n_i- \tau u^n_i}{h^{n}_{i}},\quad
 I_{10}=(u^{n+1}_{i}-u^{n}_{i})(h^{n}_{i})^2,\\
 &I_{11}=\Delta\tau Du^n_i,\quad
 I_{12}=\Delta\tau Du^n_{i+1},\quad
 I_{13}=\Delta\tau Du^n_{i-1},\quad
 I_{14}=\Delta\tau Du^n_{i-2},\\
 &I_{15}=\Delta\tau Du^{n+1}_i,\quad
 I_{16}=\Delta\tau Du^{n+1}_{i+1},\quad
 I_{17}=\Delta\tau Du^{n+1}_{i-1},\quad
 I_{18}=\Delta\tau Du^{n+1}_{i-2}.
\end{split}
\end{align}
Building the numerical scheme for the KdV equation and the lattice using these invariants guarantees that the resulting scheme is invariant under the same maximal Lie invariance group~$G$ as is the KdV equation. We first start with the discretization of~\eqref{eq:KdVComputationalCoordinates}.

It turns out that the straightforward discretization of the KdV equation in terms of the computational coordinates $(\tau,\xi)$ given by~\eqref{eq:KdVComputationalCoordinates} is already invariant under the maximal Lie invariance group $G$. We demonstrate this first for the explicit six point stencil scheme here. Indeed, the invariant finite difference expression,
\[
 I_{10}-I_{8}I_{9}\frac{I_{11}+I_{13}}{2}+\frac12\left[2\frac{I_{12}-I_{11}}{1+I_2}-2\frac{I_{11}-I_{13}}{1+I_1}+\frac{1}{I_1}\left(2\frac{I_{11}-I_{13}}{1+I_1}-2\frac{I_{13}-I_{14}}{I_1+I_3}\right)\right]=0,
\]
reads explicitly
\begin{align}\label{eq:KdVComputationalCoordinatesDiscretization6Point}
\begin{split}
 &\frac{u^{n+1}_i-u^n_i}{\Delta \tau}+ (u^n_i-\dot x_i)\frac{Du^n_i+Du^n_{i-1}}{2} + \frac{1}{2h^n_i}\bigg[\frac{2\left(Du^n_{i+1}-Du^n_{i}\right)}{h^n_{i+1}+h^n_{i}} - \frac{2\left(Du^n_{i}-Du^n_{i-1}\right)}{h^n_{i}+h^n_{i-1}}\bigg]\\
 & +\frac{1}{2h^n_{i-1}}\bigg[\frac{2\left(Du^n_{i}-Du^n_{i-1}\right)}{h^n_{i}+h^n_{i-1}} - \frac{2\left(Du^n_{i-1}-Du^n_{i-2}\right)}{h^n_{i-1}+h^n_{i-2}}\bigg]=0,
\end{split}
\end{align}
after some re-arrangements, where
\[
 \dot x_i=\frac{x^{n+1}_i-x^n_i}{\Delta \tau},
\]
denotes the \emph{grid velocity}. Correspondingly, this discretization preserves the four-dimensional maximal Lie invariance group of the KdV equation. In order to use the scheme~\eqref{eq:KdVComputationalCoordinatesDiscretization6Point} it is necessary to specify an invariant equation for the grid velocity. This will be pursued in the following subsections.

The continuous limit of scheme~\eqref{eq:KdVComputationalCoordinatesDiscretization6Point} is taken by parameterizing the spacings $h^n_i$ as a function of computational coordinates $\xi$. This implies that
\[
 h^n_i=x_\xi\Delta \xi
\]
and a Taylor series expansion of~\eqref{eq:KdVComputationalCoordinatesDiscretization6Point} gives that
\[
  u_\tau +(u-x_\tau)\frac{u_\xi}{x_\xi}+ \frac{1}{x_\xi}\left(\frac{1}{x_\xi}\left(\frac{u_\xi}{x_\xi}\right)_\xi\right)_\xi=O(
  \Delta \tau,\Delta \xi^2).
\]
Thus, as expected, the scheme~\eqref{eq:KdVComputationalCoordinatesDiscretization6Point} is of first order in time and second order in space. From the numerical point of view the scheme~\eqref{eq:KdVComputationalCoordinatesDiscretization6Point} is not advantageous as the forward in time discretization is unconditionally unstable.

A more appropriate numerical scheme can be realized on the entire ten point lattice and is given by
\begin{align*}
\begin{split}
&I_{10}-I_{8}I_{9}\frac{I_{11}+I_{13}+I_{15}+I_{17}}{4} +\frac{1}{4}\bigg[\left(2\frac{I_{16}-I_{15}}{I_{4}+I_{6}}-2\frac{I_{15}-I_{17}}{I_4+I_5}\right)
 \\ & +\frac{1}{I_5}\left(2\frac{I_{15}-I_{17}}{I_{4}+I_{5}}-2\frac{I_{17}-I_{18}}{I_5+I_7}\right)
 \\ & +\left(2\frac{I_{12}-I_{11}}{1+I_2}-2\frac{I_{11}-I_{13}}{1+I_1}\right)+\frac{1}{I_1}\left(2\frac{I_{11}-I_{13}}{1+I_1}-2\frac{I_{13}-I_{14}}{I_1+I_3}\right)\bigg]=0,
\end{split}
\end{align*}
which reads in expanded form as
\begin{align}\label{eq:KdVComputationalCoordinatesDiscretization10Point}
\begin{split}
&\frac{\hat{u}-u}{\Delta\tau}+(u^n_i-\dot x_i)\frac{Du^n_i+Du^n_{i-1}+Du^{n+1}_i+Du^{n+1}_{i-1}}{4} \\
&+\frac{1}{4h^{n+1}_i}\left[\frac{2\left(Du^{n+1}_{i+1}-Du^{n+1}_{i}\right)}{h^{n+1}_{i+1}+h^{n+1}_{i}} - \frac{2\left(Du^{n+1}_{i}-Du^{n+1}_{i-1}\right)}{h^{n+1}_{i}+h^{n+1}_{i-1}}\right]
\\&+
\frac{1}{4h^{n+1}_{i-1}}\left[\frac{2\left(Du^{n+1}_{i}-Du^{n+1}_{i-1}\right)}{h^{n+1}_{i}+h^{n+1}_{i-1}} - \frac{2\left(Du^{n+1}_{i-1}-Du^{n+1}_{i-2}\right)}{h^{n+1}_{i-1}+h^{n+1}_{i-2}}\right]
\\&+
\frac{1}{4h^n_i}\left[\frac{2\left(Du^n_{i+1}-Du^n_{i}\right)}{h^n_{i+1}+h^n_{i}} - \frac{2\left(Du^n_{i}-Du^n_{i-1}\right)}{h^n_{i}+h^n_{i-1}}\right]
\\&+
\frac{1}{4h^n_{i-1}}\left[\frac{2\left(Du^n_{i}-Du^n_{i-1}\right)}{h^n_{i}+h^n_{i-1}} - \frac{2\left(Du^n_{i-1}-Du^n_{i-2}\right)}{h^n_{i-1}+h^n_{i-2}}\right]=0.
\end{split}
\end{align}
In the continuous limit, this scheme becomes
\[
 u_\tau +(u-x_\tau)\frac{u_\xi}{x_\xi}+ \frac{1}{x_\xi}\left(\frac{1}{x_\xi}\left(\frac{u_\xi}{x_\xi}\right)_\xi\right)_\xi=O(
  \Delta \tau,\Delta \xi^2).
\]
which is still of first order in time due to the particular way the grid velocity has been discretized. Due to the use of the trapezoidal rule, the resulting scheme is conditionally stable now. The implicit six point stencil scheme is constructed in a similar fashion.

For the sake of reference we also present the standard forward in time, centered in space scheme on an orthogonal and stationary six point lattice for the KdV equation expressed in Eulerian form~\eqref{eq:KdV} here:
\begin{align*}
 &h^n_i=h=\const,\quad \Delta\tau=\Delta t,\\
 &\frac{u^{n+1}_i-u^n_i}{\Delta t}+ u^n_i\frac{u^n_{i+1}-u^n_{i-1}}{2h} + \frac{u^n_{i+2}-2u^n_{i+1}+2u^n_{i-1}-u^n_{i-2}}{2h^3}=0,
\end{align*}
It is readily checked that this discretization scheme breaks the Galilean invariance of the KdV equation while preserving invariance under both shifts and dilations. The standard, non-invariant implicit schemes on the six and ten point stencils are defined in a similar manner but not given here.

\subsection{Invariant Lagrangian discretization schemes}

In order to complete the numerical scheme~\eqref{eq:KdVComputationalCoordinatesDiscretization6Point} and~\eqref{eq:KdVComputationalCoordinatesDiscretization10Point} it is necessary to formulate an equation for the grid velocity. In the purely Lagrangian scheme one uses the discretization of the relation~\eqref{eq:RelationLagrangianEulerian}, which is
\begin{equation}\label{eq:LagrangianGridVelocity}
 \frac{x^{n+1}_i-x^n_i}{\Delta \tau}=u^n_i.
\end{equation}
That is, the grid velocity coincides with the physical velocity. It is well known that a purely Lagrangian scheme can perform poorly as there is no built-in mechanism preventing the clustering of grid points as the numerical integration proceeds~\cite{huan10By}. In the higher-dimensional case, usually mesh tangling occurs when using Lagrangian schemes.

An alternative to using~\eqref{eq:LagrangianGridVelocity} to obtain the position of the grid points on the next time level is to use \emph{adaptive moving mesh methods}. These will be shortly introduced in Section~\ref{sec:AdaptiveDiscretizationKdV}.

\subsection{Invariant evolution--projection discretization}

A possibility to make invariant Lagrangian schemes numerically competitive is to invoke them in an evolution--projection strategy~\cite{nave10Ay,seib12Ay}. The main idea is to use the invariant Lagrangian scheme introduced in the previous subsection only for a single time step and then project the solution defined on the new grid points $\{x^{n+1}_i\}$ back to the original grid $\{x^{n}_i\}$. This way, mesh movement can be effectively avoided. The projection step is in general accomplished through interpolation and the invariance of the whole solution procedure is guaranteed if the interpolation method used is invariant under the same symmetry group that has been used to construct the numerical scheme itself. This strategy has been successfully adapted for the linear heat equation and the viscous Burgers equation~\cite{bihl12Cy,bihl13By}.

We show here that polynomial interpolation of any order is invariant under the maximal Lie invariance group of the KdV equation and hence can be used in an invariant evolution--projection scheme for this equation. In the numerical results below we then choose quadratic interpolation as using it in conjunction with a second order invariant numerical scheme guarantees that the whole evolution--projection procedure is second order accurate. However, standard higher order interpolation could be used as well in invariant evolution--projection schemes for the KdV equation.

As our goal is to interpolate the solution $u^{n+1}_i$ defined at time $t^{n+1}$ back to the grid as given on time level $t^{n}$ the appropriate form of the $m$th order polynomial interpolation formula is
\begin{equation}\label{eq:GeneralPolynomialInterpolation}
 u^{n+1}(x)=\sum_{i=0}^mL_i(x)u^{n+1}_i,
\end{equation}
where
\[
 L_i(x)=\prod_{0\leqslant j\leqslant m\atop j\ne i}\frac{x-x^{n+1}_j}{x^{n+1}_i-x^{n+1}_j}
\]
are the Lagrange polynomials and $x\in[x_0^{n+1},x_m^{n+1}]$ is the point where the solution $u^{n+1}(x)$ should be interpolated. It is readily seen that the interpolation formula~\eqref{eq:GeneralPolynomialInterpolation} is invariant under space and time translations as well as under the scale symmetry of the KdV equation. Galilean invariance $(\widetilde{t^n},\widetilde{x^n_i}, \widetilde{ u^n_i})=(t^n,x^n_i+\ve t^n,u^n_i+\ve)$ is respected by~\eqref{eq:GeneralPolynomialInterpolation} too, as
\[
 \widetilde{u^{n+1}}(x)=u^{n+1}(x)+\ve=\sum_{i=0}^m\widetilde{L_i} (x)\widetilde{u^{n+1}_i} =\sum_{i=0}^m L_i(x)(u^{n+1}_i+\ve)=\left(\sum_{i=0}^m L_i(x)u^{n+1}_i\right)+\ve
\]
thus leading back to~\eqref{eq:GeneralPolynomialInterpolation}. Note that we have used here the property of the Lagrange polynomials that
\[
 \sum_{i=0}^mL_i(x)=1.
\]

Specifying the general polynomial interpolation~\eqref{eq:GeneralPolynomialInterpolation} to quadratic interpolation for the KdV equation on the ten point stencil can be done e.g.\ by setting $(x_0^{n+1},u_0^{n+1})=(x_{i-2}^{n+1},u_{i-2}^{n+1})$, $(x_1^{n+1},u_1^{n+1})=(x_{i}^{n+1},u_{i}^{n+1})$ and $(x_2^{n+1},u_2^{n+1})=(x_{i+2}^{n+1},u_{i+2}^{n+1})$. In practice, the projection step is completed by choosing the interpolating point $x\in\{x^n_{i}\}$, i.e.\ by evaluating the solution $u^{n+1}(x)$ at the location of the old grid points.

\subsection{Invariant adaptive discretization schemes}\label{sec:AdaptiveDiscretizationKdV}

Before we give the form of an invariant adaptive scheme for the KdV equation we introduce some basic background material related to adaptive numerical schemes in general. More information can be found, e.g.\ in the textbook~\cite{huan10By}.

\subsubsection{Adaptive discretization schemes}

The main idea behind moving mesh methods is to link the evolution of a mesh to the numerical solution of the discretized PDE itself. In the case of a Lagrangian scheme the new location of the grid points is determined by the solution $u$ itself only. A better criterion is usually to link the evolution of the grid points to the derivatives of $u$. This can be accomplished through the computation of \emph{equidistributing meshes}.

\begin{definition}
  Let $\rho(x)$ be a strictly positive continuous function on the interval $[a,b]$. Let $a=x_1<x_2<\cdots < x_{N-1} < x_N=b$ be a partition (i.e.\ a mesh) of this interval. The mesh is said to be equidistributing for $\rho$ on $[a,b]$ if
  \begin{equation}\label{eq:EquidistributionDefinition}
    \int_{x_1}^{x_2}\rho(x)\mathrm{d}x=\int_{x_2}^{x_3}\rho(x)\mathrm{d}x= \cdots = \int_{x_{N-1}}^{x_{N}}\rho(x)\mathrm{d}x
  \end{equation}
  holds.
\end{definition}

The function $\rho$ is called \emph{mesh density function} or monitor function. For the practical implementation it is advantageous to convert the relation~\eqref{eq:EquidistributionDefinition} into a differential equation. This is done by first using the equivalent expression
\[
 \int_{a}^{x_j}\rho(x)\mathrm{d}x=\frac{(j-1)}{N-1}\int_a^b\rho(x)\mathrm{d}x=\xi_j\int_a^b\rho(x)\mathrm{d}x,
\]
where $\xi_j$, $j=1,\dots,N$, is the discrete computational coordinate. By definition, $\xi_j\in[0,1]$.

Regarding $x$ as a function of the computational coordinate, i.e.\ $x_j=x(\xi_j)$, in the continuous limit the above integral equation becomes
\[
 \int_{a}^{x(\xi)}\rho(x)\mathrm{d}x=\xi\int_a^b\rho(x)\mathrm{d}x,
\]
which holds for all $\xi\in[0,1]$. Differentiating this equation twice with respect to $\xi$ leads to
\begin{equation}\label{eq:EquidistributionPrinciple}
 (\rho(x)x_\xi)_\xi=0,
\end{equation}
which is the differential form of the equidistribution principle when subjected to the boundary conditions $x(0)=a$ and $x(1)=b$.

So as to complete the description of a numerical scheme upon using the equidistribution principle in its differential form~\eqref{eq:EquidistributionPrinciple} one needs to specify the mesh density function~$\rho$. A classical choice is the arc-length type function
\[
 \rho=\sqrt{1+\alpha u_x^2},
\]
where $\alpha\in\mathbb{R}$ is a constant parameter governing the strength of the adaptation. Other monitor functions, such as built around the curvature of $u$ are used as well.

\subsubsection{Invariant adaptive scheme for the KdV equation}

In order to complete the invariant numerical scheme for the KdV equation one has to discretize the differential form of the equidistribution principle~\eqref{eq:EquidistributionPrinciple} in an invariant way. As the missing ingredient in the grid velocity $\dot x_i$ is $x^{n+1}_i$, we discretize~\eqref{eq:EquidistributionPrinciple} on the time layer $t^{n+1}$. This is done upon composing a discretization of~\eqref{eq:EquidistributionPrinciple} out of the difference invariants for the KdV equation~\eqref{eq:DifferenceInvariants10points}. A possible discretization using the arc-length type mesh density function is:
\[
 \frac{\rho^n_{i+1}+\rho^n_{i}}{2}I_{11}- \frac{\rho^n_{i}+\rho^n_{i-1}}{2}\frac{I_{13}}{I_1}=0,
\]
where
\[
 \rho_{i+1}=\sqrt{1+\alpha I_{11}^2},\quad \rho_{i}=\sqrt{1+\alpha I_{12}^2},\quad \rho_{i-1}=\sqrt{1+\alpha I_{13}^2},
\]
or, explicitly,
\begin{equation}\label{eq:DiscretizedEquidistribution}
 \frac{\rho^n_{i+1}+\rho^n_{i}}{2}(x^{n+1}_{i+1}-x^{n+1}_{i})- \frac{\rho^n_{i}+\rho^n_{i-1}}{2}(x^{n+1}_{i}-x^{n+1}_{i-1})=0,
\end{equation}
where
\begin{equation}\label{eq:InvariantMeshDensityFunction}
 \rho^n_i=\sqrt{1+\alpha\left(\Delta\tau\frac{u^n_{i+1}-u^n_{i}}{x^n_{i+1}-x^n_{i}}\right)^2}.
\end{equation}

\subsection{Momentum preserving invariant discretization}

It is well-known that the KdV equation admits infinitely many conservation laws, see e.g.~\cite{olve86Ay} for a discussion. Numerically preserving conservation laws of partial differential equations is generally a nontrivial problem that belongs to the realm of geometric numerical integration. More information on this field can be found in the books~\cite{hair06Ay,leim04Ay}. The problem of finding finite difference discretizations for the KdV equation that preserve sub-sets of the infinite span of conservation laws is a complicated problem that will not be investigated here. We are only concerned with finding invariant discretization schemes that also preserve linear momentum
\[
 \mathcal M=\int u\,\mathrm{d}x.
\]
This conservation law is associated with expressing the KdV equation itself in conserved form
\[
 \mathrm{D}_t u + \mathrm{D}_x\left(\frac12u^2+u_{xx}\right)=0.
\]
It is possible to preserve the above conserved form also on a moving mesh, which as we have seen above is a basic requirement for preserving Galilean invariance. In particular, the following discretization is invariant under the maximal Lie invariance group of the KdV equation and momentum-preserving:
\begin{align}\label{eq:ConservativeDiscretizationKdV}
\begin{split}
 &\frac{(h^{n+1}_i+h^{n+1}_{i-1})u^{n+1}_i-(h^{n}_i+h^{n}_{i-1})u^{n}_i}{\Delta\tau}-\left(\frac{x^{n+1}_{i+1}-x^{n}_{i+1}}{\Delta\tau}u^n_{i+1} -\frac{x^{n+1}_{i-1}-x^{n}_{i-1}}{\Delta\tau}u^n_{i-1}\right)\\
 &+\frac12((u^n_{i+1})^2-(u^n_{i-1})^2)+\left[\frac{2(Du^n_{i+1}-Du^n_{i})}{h^n_{i+1}+h^n_{i}}-\frac{2(Du^n_{i-1}-Du^n_{i-2})}{h^n_{i-1}+h^n_{i-2}}\right]=0.
\end{split}
\end{align}
The associated continuous expression to this discretization is
\[
 (x_\xi u)_\tau+\left(\frac12u^2+\left(\frac{1}{x_\xi}\left(\frac{u_\xi}{x_\xi}\right)_\xi\right)_\xi-u x_\tau\right)_\xi=0,
\]
which is of conserved form in the computational coordinates. It thus discretely conserves momentum $\mathcal M$.

Let us now show that~\eqref{eq:ConservativeDiscretizationKdV} also preserves all the Lie symmetries as admitted by the KdV equation. One way of showing this would be to express~\eqref{eq:ConservativeDiscretizationKdV} in terms of the difference invariants~\eqref{eq:DifferenceInvariants10points}. However, due to the particular form of~\eqref{eq:ConservativeDiscretizationKdV} a direct expression in terms of difference invariants would be cumbersome. It is much easier to verify invariance directly by transforming the scheme~\eqref{eq:ConservativeDiscretizationKdV} under the action of the symmetry group of the KdV equation.

It is obvious that the discretization~\eqref{eq:ConservativeDiscretizationKdV} is invariant under shifts in space and time as well as under scale transformations. It thus only remains to show invariance under Galilean transformations $(t^n,x^n_i,u^n_i)\mapsto(t^n,x^n_i+\ve t^n,u^n_i+\ve)$. We proceed term by term:
\begin{align*}
 &\frac{(\widetilde{h^{n+1}_i}+\widetilde{h^{n+1}_{i-1}})\widetilde{u^{n+1}_i}-(\widetilde{h^{n}_i}+ \widetilde{h^{n}_{i-1}})\widetilde{u^{n}_i}}{\widetilde{\Delta \tau}}=\frac{(h^{n+1}_i+h^{n+1}_{i-1})u^{n+1}_i-(h^{n}_i+h^{n}_{i-1})u^{n}_i}{\Delta\tau}\\
 &\qquad\qquad+\ve\left(\frac{x^{n+1}_{i+1}-x^{n}_{i+1}}{\Delta\tau} -\frac{x^{n+1}_{i-1}-x^{n}_{i-1}}{\Delta\tau}\right),\\
 &\left(\frac{\widetilde{x^{n+1}_{i+1}}-\widetilde{x^{n}_{i+1}}}{\widetilde{\Delta \tau}}\widetilde{u^n_{i+1}} -\frac{\widetilde{x^{n+1}_{i-1}}-\widetilde{x^{n}_{i-1}}}{\widetilde{\Delta\tau}}\widetilde{u^n_{i-1}}\right)=\left(\frac{x^{n+1}_{i+1}-x^{n}_{i+1}}{\Delta\tau}u^n_{i+1} -\frac{x^{n+1}_{i-1}-x^{n}_{i-1}}{\Delta\tau}u^n_{i-1}\right)\\
 &\qquad\qquad+\ve(u^n_{i+1}-u^n_{i-1})+ \ve\left(\frac{x^{n+1}_{i+1}-x^{n}_{i+1}}{\Delta\tau} -\frac{x^{n+1}_{i-1}-x^{n}_{i-1}}{\Delta\tau}\right),\\
 & \frac12((\widetilde{u^n_{i+1}})^2-(\widetilde{u^n_{i-1}})^2)=\frac12((u^n_{i+1})^2-(u^n_{i-1})^2)+\ve(u^n_{i+1}-u^n_{i-1}),\\
 &\left[\frac{2(\widetilde{Du^n_{i+1}}-\widetilde{Du^n_{i})}}{\widetilde{h^n_{i+1}}+ \widetilde{h^n_{i}}}-\frac{2(\widetilde{Du^n_{i-1}}-\widetilde{Du^n_{i-2}})}{\widetilde{h^n_{i-1}} +\widetilde{h^n_{i-2}}}\right]=\\ &\qquad\qquad\left[\frac{2(Du^n_{i+1}-Du^n_{i})}{h^n_{i+1}+h^n_{i}}-\frac{2(Du^n_{i-1}-Du^n_{i-2})}{h^n_{i-1}+h^n_{i-2}}\right]
\end{align*}
Substituting into the transformed form of equation~\eqref{eq:ConservativeDiscretizationKdV} proves Galilean invariance.

As it stands, the momentum preserving invariant scheme~\eqref{eq:ConservativeDiscretizationKdV} still needs to be completed by adapting an appropriate strategy to obtain the new mesh $\{x^{n+1}_i\}$. Here, the same strategies as proposed above for the case of the non-conservative invariant scheme~\eqref{eq:KdVComputationalCoordinatesDiscretization10Point} can be applied. These strategies lead to \emph{invariant momentum-preserving Lagrangian}, \emph{evolution--projection} and \emph{adaptive schemes}, respectively.

\subsection{Exact discretization}\label{sec:ExactDiscretization}

An interesting question on the behavior of numerical schemes is whether they are able to reproduce exact solutions of the original differential equation exact, i.e.\ without numerical error.

Among all the exact solutions given in Section~\ref{sec:SymmetryReduction}, the only solutions that are exact for all schemes reported in Section~\ref{sec:DiscretizationKdV} is the constant solution~\eqref{eq:ConstantSolution}. In addition, the Galilean invariant solution~\eqref{eq:GalileiInvariantSolution} is an exact solution for the invariant Lagrangian schemes~\eqref{eq:KdVComputationalCoordinatesDiscretization6Point} and~\eqref{eq:KdVComputationalCoordinatesDiscretization10Point} using~\eqref{eq:LagrangianGridVelocity} which is readily verified directly. Below we verify numerically that this solution is also exact for the invariant evolution--projection scheme and the invariant momentum preserving scheme.

\section{Numerical results}\label{sec:ComputationalResultsKdV}

In this section we collect the numerical results obtained using the various schemes proposed in the previous section. Our purpose is not to do a technical optimization of every scheme but to rather demonstrate the feasibility of implementing invariant discretization schemes as well as the resulting physical implications.

For the invariant adaptive scheme, we use the discretization~\eqref{eq:DiscretizedEquidistribution} of the equidistribution principle with the invariant mesh density function~\eqref{eq:InvariantMeshDensityFunction}. To compare the invariant adaptive scheme against a non-invariant adaptive one we also use the mesh density function $\rho=\sqrt{1+\alpha u_{xx}^2}$, discretized by
\begin{equation}\label{eq:NoninvariantMeshDensityFunction}
\rho^n_i {}_\text{non-inv}=\sqrt{1+\alpha\left(\Delta\tau \frac{2\frac{u^n_{i+2}-u^n_{i}}{x^n_{i+2}-x^n_{i}}-2\frac{u^n_{i+1}-u^n_{i-1}}{x^n_{i+1}-x^n_{i-1}}}{x^n_{i+2}-x^n_{i}+x^n_{i+1}-x^n_{i-2}}\right)^2}.
\end{equation}
in conjunction with~\eqref{eq:DiscretizedEquidistribution}. Similar mesh density functions are also used in adaptive numerical schemes, see e.g.~\cite{huan10By}. In the present case, using~\eqref{eq:NoninvariantMeshDensityFunction} breaks the scale invariance in the discretization of the KdV equation. The resulting scheme therefore serves as reference for a non-invariant adaptive scheme.

Note that for the sake of brevity we abbreviate the standard notation $a\cdot 10^n$ in the tables and figure legends below by a$e$n.

\subsection{Decaying cosine evolution}

Before we use the exact solutions computed in Section~\ref{sec:SymmetryReduction} as benchmark tests, we reproduce the classical results obtained by Zabusky and Kruskal in 1965~\cite{zabu65Ay} of a wave decaying into solitons. For this experiment, Zabusky and Kruskal used the following form of the KdV equation
\[
 u_t+u u_x+ \delta^2  u_{xxx} = 0,
\]
where $\delta=0.022$. The initial condition used was $u=\cos(\pi x)$ on a periodic domain of length $L=2$. Zabusky and Kruskal observed the formation of eight solitons at time $t=3.6/\pi$.

A main problem reproducing this result with the invariant Lagrangian schemes~\eqref{eq:KdVComputationalCoordinatesDiscretization6Point} and~\eqref{eq:KdVComputationalCoordinatesDiscretization10Point} using~\eqref{eq:LagrangianGridVelocity} is that mesh tangling occurs before the final integration time $t=3.6/\pi$. In turn, using the invariant Lagrangian scheme only in the framework of the invariant evolution--projection method allows us to arrive at a solution at the final integration time.

\begin{figure}[!ht]
 \centering
 \includegraphics[scale=0.75]{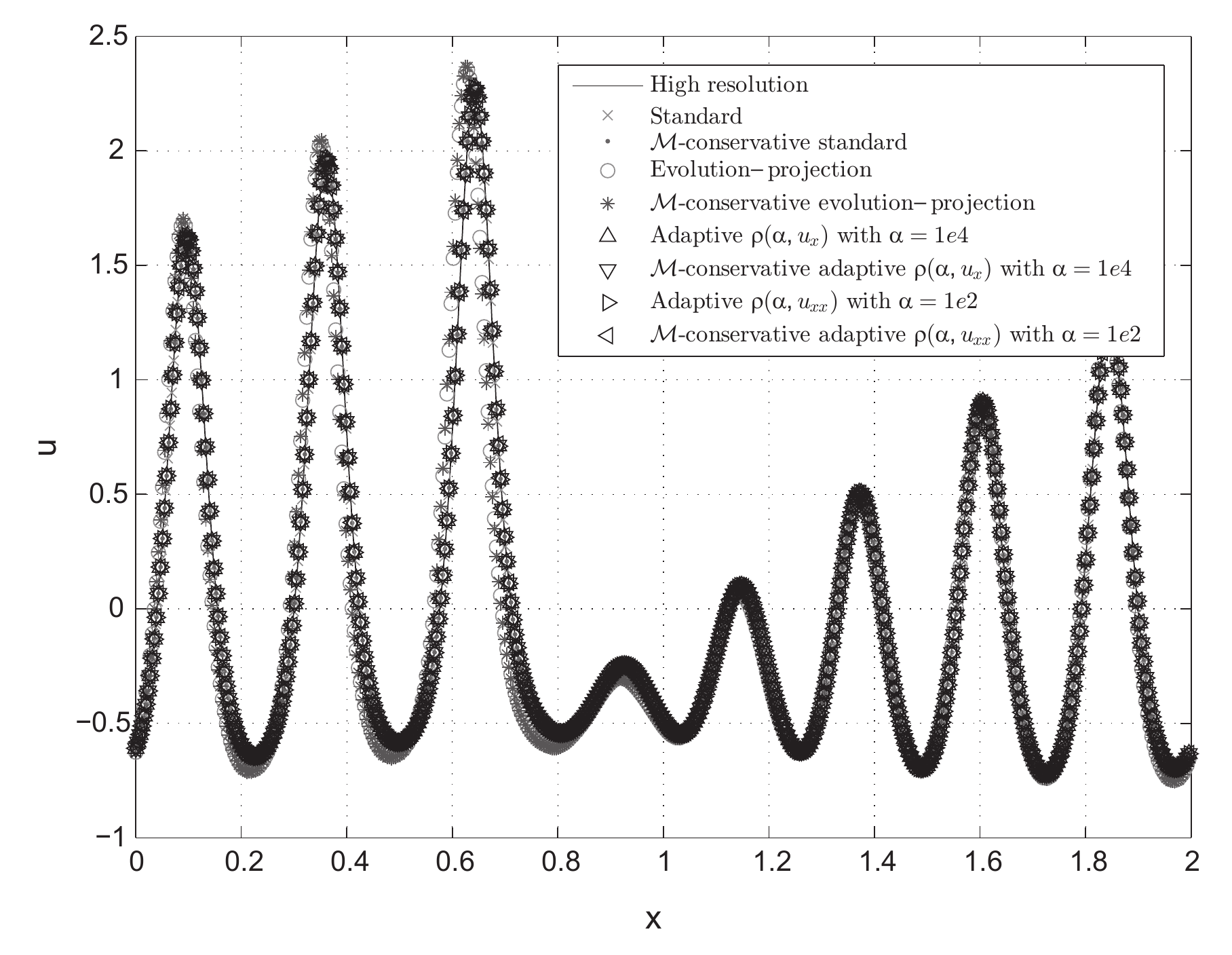}
 \caption{Numerical solution of the Zabusky--Kruskal decaying into soliton problem. The following schemes were tested on the ten point stencil, using $N=512$ mesh points except for the high resolution reference run (solid line) for which $N=2048$ points were used: Non-invariant standard finite differences (crosses), non-invariant momentum conservative (dots), invariant evolution--projection (open circles), invariant evolution--projection momentum conservative (stars), invariant adaptive with monitor function $\rho=\sqrt{1+10^{4}u_x^2}$ (upward pointing triangles), invariant adaptive momentum-preserving with monitor function $\rho=\sqrt{1+10^{4}u_x^2}$ (downward pointing triangles), non-invariant adaptive with monitor function $\rho=\sqrt{1+10^{2}u_{xx}^2}$ (rightward pointing triangles), non-invariant adaptive momentum-preserving with monitor function $\rho=\sqrt{1+10^{2}u_{xx}^2}$ (leftward pointing triangles). See Table~\ref{tab:cos_into_sol} for a quantification of these numerical experiments that are visually practically indistinguishable.} \label{fig:cosine_into}
\end{figure}

All the other schemes presented above are able to compute this test problem. The results of these integrations are shown in Fig.~\ref{fig:cosine_into}. From this figure it can be seen that all schemes are capable of capturing the decay into solitons as originally presented in~\cite{zabu65Ay}. It can also be seen that the two evolution--projection schemes show a slight lag for the first four solitons when compared to the high resolution solution. The other schemes lie visually very close to this high resolution solution.

To quantify these findings, in Table~\ref{tab:cos_into_sol} we present the root-mean-square error (RMSE) for the various schemes tested, using the high resolution finite difference solution as reference. The RMSE is defined~by
\[
\mathrm{RMSE} = \sqrt{\sum_{i=1}^{N}\frac{(u_{\rm num_i}-u_{\rm exact_i})^2}{N}}.
\]
where in place of the exact solution, $u_{\rm exact}$, the high resolution numerical solution is used.

It can be seen from this table that the evolution--projection schemes have indeed errors larger by a factor of ten than the other schemes tested, which all give quite comparable errors. A possible explanation for this increase of error is that the interpolation used does not accurately take into account the rapid change in the first derivatives of the numerical solution. Using higher order interpolation incorporating derivative information, such as Hermite interpolation, could help reduce this phase error in the evolution--projection scheme, see also~\cite{nave10Ay,seib12Ay}.

\begin{table}[!ht]
\begin{center}
\begin{tabular}{ |c | c | }
\hline
	Scheme & RMSE \\ \hline\hline
	Non-invariant standard & 0.0138    \\ \hline
	Non-invariant standard $\mathcal M$-cons & 0.0139 \\ \hline
	Invariant evolution--projection & 0.189    \\ \hline
	Invariant evolution--projection $\mathcal M$-cons & 0.202   \\ \hline
	Invariant adaptive ($\rho(\alpha,u_x)$ with $\alpha=1e4$) & 0.0142    \\ \hline
	Invariant adaptive $\mathcal M$-cons ($\rho(\alpha,u_x)$ with $\alpha=1e4$) & 0.0139    \\ \hline
	Non-invariant adaptive ($\rho(\alpha,u_{xx})$ with $\alpha=1e2$) & 0.0144   \\ \hline
	Non-invariant adaptive $\mathcal M$-cons ($\rho(\alpha,u_{xx})$ with $\alpha=1e2$) & 0.0138   \\ \hline
\end{tabular}
\end{center}
\caption{Numerical errors for the Zabusky--Kruskal problem. The RMSE is based on a high resolution integration using $N=2048$ mesh points and a time step $\Delta t=3.125\cdot 10^{-7}$ in the non-invariant standard numerical scheme for the KdV equation. All other schemes use $N=512$ mesh points with time step $\Delta t=5\cdot 10^{-6}$.}\label{tab:cos_into_sol}
\end{table}

\subsection{Exact algebraic solution}

As was discussed in Section~\ref{sec:ExactDiscretization}, the invariant Lagrangian schemes~\eqref{eq:KdVComputationalCoordinatesDiscretization6Point} and~\eqref{eq:KdVComputationalCoordinatesDiscretization10Point} using~\eqref{eq:LagrangianGridVelocity} are exact for the Galilean invariant solution~\eqref{eq:GalileiInvariantSolution}. We verify this by numerically computing this solution and calculating the $l_\infty$-norm and the RMSE. The $l_\infty$-norm is the maximum absolute difference between the numerical solution $u_{\rm num}$ and the exact analytical solution $u_{\rm exact}$ calculated at the discrete mesh points.

The results as seen in Table~\ref{fig:exact} show that we achieve machine precision (i.e.\ the errors come only from rounding) with the different invariant schemes introduced in Section~\ref{sec:DiscretizationKdV} but do not get comparable accuracy with the standard schemes. Table~\ref{fig:exact} also highlights that the evolution--projection method (both invariant and invariant momentum conserving) reproduces the exact solution up to machine precision as well.

\begin{table}[!ht]
\begin{center}
\begin{tabular}{ | c |c | c | }
\hline
	 Scheme  & $l_\infty$-norm  & RMSE  \\ \hline\hline
	 Non-invariant standard  &  6.76e-6  &  2.39e-6  \\ \hline
	 Non-invariant standard $\mathcal M$-cons &  7.77e-6 &  3.30e-6 \\ \hline
     Invariant Lagrangian &  4.73e-13  &  2.02e-13  \\ \hline
	 Invariant evolution--projection  & 2.13e-14 & 7.93e-15 \\ \hline
	 Invariant Lagrangian $\mathcal M$-cons &  9.73e-13  &  4.02e-14 \\ \hline
	 Invariant evolution--projection $\mathcal M$-cons & 5.15e-14 & 1.31e-14 \\ \hline
\end{tabular}
\end{center}
\caption{Comparison of errors for the various ten point schemes to reproduce the exact solution~\eqref{eq:GalileiInvariantSolution} evaluated at $t=2$. All schemes use $N=35$ mesh points on the domain $[0,20]$ and time steps of $\Delta\tau=0.001$. The starting time of the integrations is $t_0=1$.}
\label{fig:exact}
\end{table}

This solution, being monotonously increasing, is one of the few where the Lagrangian moving mesh points cause no instability over a longer period of time. No interpolation or adaptation is therefore needed to get an exact solution at any time. We should also stress that for this simple solution the adaptive schemes would coincide with the standard scheme as $u_x=1/t$ and $u_{xx}=0$ thus reducing the discretized equidistribution principle~\eqref{eq:DiscretizedEquidistribution} for both the invariant and non-invariant mesh density functions~\eqref{eq:InvariantMeshDensityFunction} and~\eqref{eq:NoninvariantMeshDensityFunction} to $x^{n+1}_{i+1}-x^{n+1}_{i}=x^{n+1}_{i}-x^{n+1}_{i-1}$.

While integrating such a simple function is trivial, this example shows the compatibility of preserving symmetries and obtaining exact discrete solutions.

\subsection{Cnoidal wave and soliton solution}
	
For any numerical scheme, one important test is to verify consistency and the order of convergence. To verify the order of the numerical schemes proposed in this paper, we take cnoidal wave periodical solution of the form
\begin{equation}\label{eq:CnoidalWaveNumerics}
u=(a-b) cn^2(\omega(x+bt),k)
\end{equation}
where $a=3.332$, $b=-0.784$, $c=-2.548$, $	k=\sqrt{\frac{a-b}{a-c}}=\sqrt{0.7}$ and $\omega=\sqrt{\frac{a-c}{12}}=0.7$

We then vary the total number of mesh points $n\in \{16,24,32,38\}$ and measure for each associated numerical experiment the error characterized by the $l_\infty$-norm of the difference between the numerical and the discrete analytical solutions. A linear regression of $log(\text{n})$ vs.\ $log(\text{error})$ gives a slope corresponding to the order of convergence in $\mathcal{O}(n^p)$. All our ten point schemes should theoretically converge as $\mathcal{O}(n^{-2})$ and we notice in Table~\ref{fig:cn_convergence} that this is numerically effectively the case.
	
\begin{table}[!ht]
\begin{center}
\begin{tabular}{ |c | c | c | c }
\hline
	Scheme & $p\ \textup{in}\ \mathcal O(N^p)$ \\ \hline\hline
	Non-invariant standard & -2.00    \\ \hline
	Non-invariant standard $\mathcal M$-cons & -2.05 \\ \hline
	Invariant Lagrangian & -2.13    \\ \hline
	Invariant Lagrangian $\mathcal M$-cons & -2.11  \\ \hline
	Invariant evolution--projection & -1.91    \\ \hline
	Invariant evolution--projection $\mathcal M$-cons & -2.04    \\ \hline
	Invariant adaptive ($\rho(\alpha,u_x)$ with $\alpha=5e6$) & -2.05    \\ \hline
	Invariant adaptive $\mathcal M$-cons ($\rho(\alpha,u_x)$ with $\alpha=5e6$) & -2.05    \\ \hline
	Non-invariant adaptive ($\rho(\alpha,u_{xx})$ with $\alpha=1e6$) & -2.00    \\ \hline
	Non-invariant adaptive $\mathcal M$-cons ($\rho(\alpha,u_{xx})$ with $\alpha=1e6$) & -2.02  \\ \hline
\end{tabular}
\end{center}
\caption{Convergence tests are done for the cnoidal solution over one spatial period at time $t=0.2$ with time step $\Delta t=10^{-4}$. All schemes use the ten point lattice. The integrations are done using $N=\{16,24,32,48\}$ points. We confirm that all schemes converge as $\mathcal{O}(N^{-2})$ in the $l_\infty$-norm and are therefore consistent.}\label{fig:cn_convergence}
\end{table}

To assess not only the order of the numerical schemes but also the absolute approximation errors in Table~\ref{tab:cn_and_soli} we present the RMSE comparing the numerical solution against the exact cnoidal wave solution as given in~\eqref{eq:CnoidalWaveNumerics}. As a second example, we also compare against the soliton solution
\begin{equation}
u=\frac{3\nu}{\cosh^2{(\frac{1}{2}\sqrt{\nu}(x-\nu t))}}
\end{equation}
with $\nu=7$. In addition to the approximation error we also monitor the change in momentum~$\Delta \mathcal M$ over the integration period.

\begin{table}[!ht]
\begin{center}
\begin{tabular}{ | c | c | c |c|c| }
\hline
	&\multicolumn{2}{c|}{Cnoidal wave} &  \multicolumn{2}{c|}{Soliton}    \\ \hline
	Scheme &  RMSE &$\Delta \mathcal M$ & RMSE & $\Delta \mathcal M$ \\ \hline \hline
Non-inv standard & 3.98e-3 &1.96e-14 & 9.56e-2 & 3.20e-14 \\ \hline
Non-inv standard $\mathcal M$-cons & 1.52e-3 &2.66e-14 & 3.38e-2 & 7.11e-14  \\ \hline
Inv 5 point explicit Lagrangian & 4.59e-2 & 8.07e-3 & 0.439 & 0.367  \\ \hline
Inv Lagrangian & 7.69e-3 &1.26e-3 & 0.346  & 0.166\\ \hline
Inv Lagrangian $\mathcal M$-cons & 9.91e-3 &2.31e-14 & 0.436 & 3.91e-14 \\ \hline
Inv	evolution--projection & 4.93e-3 &8.19e-4 & 0.288 & 7.78e-2 \\ \hline
Inv evolution--projection $\mathcal M$-cons &5.58e-3 &7.59e-4 & 0.327 & 4.08e-2 \\ \hline \hline
Non-inv adaptive ($\rho(\alpha,u_{xx})$, $\alpha=1e6$) & 3.92e-3 & 2.71e-6&  ---&--- \\ \hline
Non-inv adaptive $\mathcal M$-cons ($\rho(\alpha,u_{xx})$, $\alpha=1e6$) & 1.57e-3 & 2.31e-14 &  ---&--- \\ \hline
Inv adaptive ($\rho(\alpha,u_x)$, $\alpha=5e6$) & 3.99e-3& 1.54e-5 & ---& --- \\ \hline
Inv adaptive $\mathcal M$-cons ($\rho(\alpha,u_x)$, $\alpha=5e6$) & 1.48e-3& 3.38e-14 & ---&---  \\ \hline
Non-inv adaptive ($\rho(\alpha,u_{xx})$, $\alpha=1e4$) & --- &---& 9.49e-2 &9.12e-4  \\ \hline
Non-inv adaptive $\mathcal M$-cons ($\rho(\alpha,u_{xx})$, $\alpha=1e4$) & --- &---& 2.94e-2 &4.97e-14  \\ \hline
Inv adaptive ($\rho(\alpha,u_x)$, $\alpha=1e4$) & --- &---& 9.28e-2& 5.74e-4 \\ \hline
Inv adaptive $\mathcal M$-cons ($\rho(\alpha,u_x)$, $\alpha=1e4$) & --- &---& 0.682 &3.55e-14  \\ \hline

\end{tabular}
\end{center}
\caption{Errors of different schemes for the cnoidal wave and soliton solutions. All schemes use the ten point lattice unless otherwise stated. Time steps are always $\Delta t=10^{-4}$. The cnoidal wave is integrated over one period up to $t=0.2$ while the soliton is computed up to $t=0.05$ on the domain $[-4,4]$. The short integration time is to allow using the purely Lagrangian method. For both integration $N=48$ total mesh points are used. The projection method is parabolic interpolation. The suitable adaptation parameter depends on both the form of the monitor function and the initial conditions.}
\label{tab:cn_and_soli}
\end{table}

By computing the RMSE of the different invariant and non-invariant schemes, in Table~\ref{tab:cn_and_soli} we are able to affirm that invariant and non-invariant schemes give roughly the same approximation errors. The invariant adaptive and non-invariant adaptive scheme give comparable accuracy while the standard scheme is slightly better than the Lagrangian scheme. We confirm that the invariant ten point scheme gives better accuracy than the invariant explicit scheme on the five point lattice as expected. The basic projection method using parabolic interpolation helps to reduce the error and allows using longer integration times. Optimizing the adaptation parameter $\alpha$ is possible as well (see~\cite{huan10By}) and could lead to error improvements. This will however not be pursued here.

\subsection{Double soliton solution and Galilean invariance}

The above numerical experiments show that in terms of accuracy the invariant and the non-invariant schemes are mostly comparable (except for the exact solution~\eqref{eq:GalileiInvariantSolution}). Still, from the physical point of view, the additional advantage of the invariant schemes over the standard ones is the preservation of Galilean invariance. In particular, Galilean invariance in a numerical scheme implies that applying a boost to any solution does not change the discrete numerical solution. Hence, the numerical solutions can be obtained in any constantly moving reference frame. This can be an important property in practical applications, see e.g.~\cite{bihl13By} and references therein for applications of this property to hydrodynamics.

To numerically verify Galilean invariance in the proposed invariant schemes, we integrate the double soliton solution over a short period of time and apply a boost to the invariant and the non-invariant schemes. The following form of the double soliton solution is used:
\begin{align*}
&\tilde{u}(t,\tilde{x})=12\frac{\partial^2}{\partial x^2}\ln(1+B_1e^{iQ_1}+B_2e^{iQ_2}+AB_1B_2e^{i(Q_1+Q_2)})+c,\\
&Q_1=a_1\tilde{x}-a_1^3t,\quad a_2\tilde{x}-a_2^3t,\quad A=\left(\frac{a_1-a_2}{a_1+a_2}\right)^2
\end{align*}
where $a_1=-2i$, $a_2=-i$, $B_1=10000$, $B_2=1$, \{$\tilde{u},\tilde{x}=x-ct$\} belong to the moving reference frame and $c$ is the speed of the moving reference frame.

Two sets of numerical experiments are carried out for each scheme. One in a resting reference frame, i.e.\ $c=0$ and one in a constantly moving reference frame, $c\ne0$. After the end of each integration both solutions are compared to each other. Galilean invariance implies that both solutions must coincide up to machine precision.

\begin{figure}[!ht]
 \centering
 \includegraphics[scale=0.75]{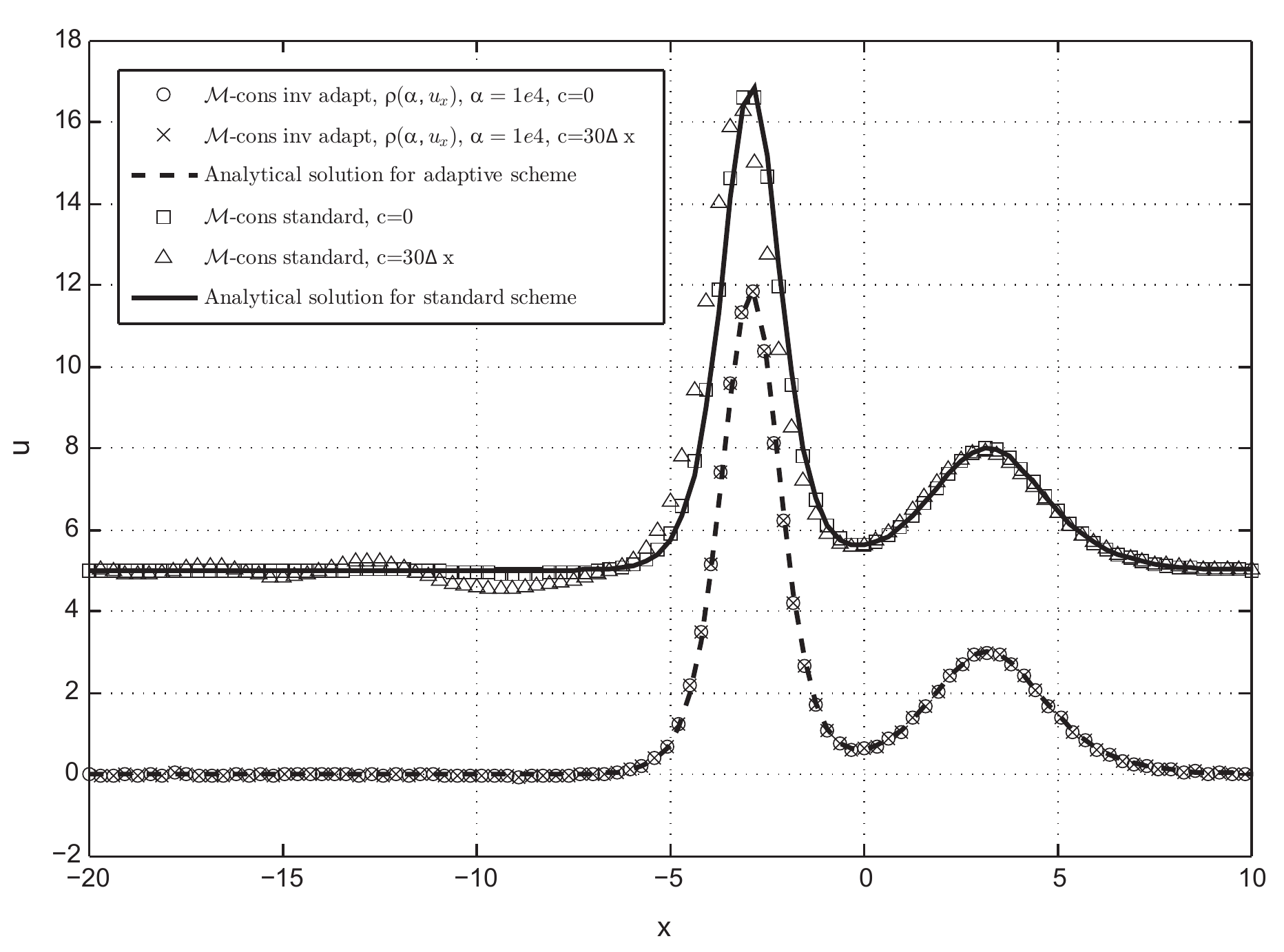}
 \caption{Double soliton solution at time $t=0.1$ computed using the non-invariant and invariant adaptive momentum-conservative schemes. Time step used is $\Delta t=10^{-4}$ with $N=128$ grid points. The non-invariant solutions are shifted with respect to the invariant solutions for the sake of comparison. While the non-invariant scheme in a resting reference frame (open squares) approximates closely the exact solution (dashed line), using this scheme in a constantly moving reference frame (triangles) leads to large deviations from the true solution. For the invariant scheme, both the solution in the resting reference frame (open circles) and in a constantly moving reference frame (crosses) are in good accordance with the exact solution (solid line).}\label{fig:2soliton}
\end{figure}

By increasing the strength of the boost and by computing the RMSE, we observe an increase in the error for the non-invariant momentum-conserving scheme while the invariant adaptive and momentum-conserving scheme is largely unaffected, see Table~\ref{fig:table_2soliton} for quantification. The Galilean boosted solution for the invariant scheme in Fig.~\ref{fig:2soliton} is identical to its equivalent in the resting reference frame and visually confirms the Galilean invariance of this scheme, a major physical property lost when using standard non-invariant schemes. For the other invariant and non-invariant schemes the results are essentially the same and are hence not presented here.

\begin{table}[!ht]
\begin{center}
\begin{tabular}{|c|c|c|}
 \hline
 	 & \multicolumn{2}{c|}{RMSE compared to c=0 solution}  \\ \hline
 	 		$c/\Delta x$ &Non-inv standard $\mathcal M$-cons & Inv adaptive $\mathcal M$-cons ($\rho(\alpha,u_x)$, $\alpha=1e4$)  \\
\hline -10 & 2.14e-1 & 1.82e-12  \\
\hline -1 & 2.29e-2 & 1.07e-12 \\
\hline 0 & 0 & 0  \\
\hline 1 & 2.19e-2 & 8.76e-13  \\
\hline 5 & 0.103 & 3.15e-12 \\
\hline 10 & 0.202 & 1.86e-12  \\
\hline 30 & 0.564 & 1.03e-12 \\
\hline
\end{tabular}
\end{center}
\caption{RMSE comparing the resting reference solution ($c=0$) to a constantly moving solution ($c\ne0$) for the non-invariant momentum-preserving scheme (left) and the invariant adaptive momentum-preserving scheme (right). Integrations were done up to $t=1$ using the time step $\Delta t=10^{-3}$ and $N=128$ points. It can be seen that varying the speed $c$ of the reference frame leads to significantly different solutions for the standard scheme as measured through the RMSE while for the invariant scheme the RMSE stays approximately constant and is due to rounding only.}\label{fig:table_2soliton}
\end{table}

\section{Conclusions}\label{sec:ConclusionKdV}

In this paper we have constructed invariant numerical schemes for the Korteweg--de Vries equation. While some invariant numerical schemes have already been constructed for this equation in the past~\cite{doro94Ay,doro91Ay}, to the best of our knowledge this is the first time that actual numerical experiments have been carried out for the KdV equation using such schemes. We found that these existing schemes, all Lagrangian in nature, can develop tangling meshes and hence may not allow integration beyond some fixed time limit. A remedy for these schemes is provided through invoking them in an evolution--projection framework. As shown for several test cases, these evolution--projection schemes can produce numerical solutions for the KdV equation without being restricted by the development of mesh problems.

In addition, we have proposed several Eulerian numerical schemes most notably by using ideas of adaptive moving mesh methods. These schemes are attractive in that they link the required moving meshes (to preserve Galilean invariance) to the development of pronounced features of the numerical solution. Hence, such schemes are capable of tracking developing shots, blow-ups etc. Furthermore, we have shown that it is possible to develop invariant numerical schemes that also preserve momentum.

In terms of accuracy, we have found that the invariant and non-invariant schemes are comparable. This is in striking contrast to symmetry-preserving integrators for ODES, where invariant schemes can perform significantly better than their non-invariant counterparts, especially for solutions with singularities~\cite{bour06Ay,bour08Ay,rebe09Ay}. For the KdV equation, one possible explanation for this discrepancy is that the maximal Lie invariance group is of rather simple structure, with three of the four admitted one-parameter symmetry transformations (shifts in space and time as well as dilations) already preserved by standard numerical schemes for this equation. Hence, the only difference between the non-invariant and invariant schemes for the KdV equation is whether Galilean invariance is admitted or not.

One manifestation of this difference is the achieved accuracy for the Galilean invariant solution~\eqref{eq:GalileiInvariantSolution}. While for the invariant schemes this solution is also exact, this is not the case for the standard schemes. This explains the significantly better accuracy of the invariant schemes compared to the non-invariant ones.

While getting better numerical schemes is a main motivation for research in numerical analysis, it is also important to develop schemes that accurately capture the essential properties of a physical model. The KdV equation, like most other classical equations of classical hydrodynamics is invariant under the Galilean group. It is hence of primary importance to devise schemes that are able of preserving the Galilean group numerically. As was shown in the original work by Dorodnitsyn~\cite{doro11Ay} this is only possible on moving discretization meshes. With Lagrangian schemes not being a numerically competitive option, the here proposed discretizations in computational coordinates and combination with proven adaptation methods are a viable route to numerically preserve this important Lie group.

\section*{Acknowledgements}

This research was supported by the Austrian Science Fund (FWF), project J3182--N13 (AB). AB is a recipient of an APART Fellowship of the Austrian Academy of Sciences. The research of PW was partially supported by NSERC of Canada.

{\footnotesize\setlength{\itemsep}{0ex}
\bibliography{bihloKdV}
}

\end{document}